\documentclass{elsart}

\newcommand{\1}{{\Delta T }}

\newcommand{\beq}{\begin{equation}}
\newcommand{\eeq}{\end{equation}}
\newcommand{\lab}{\label}

\newcommand{\bfxi}{\mbox{\boldmath $\xi$}}
\newcommand{\bfeta}{\mbox{\boldmath $\eta$}}
\newcommand{\bfal}{\mbox{\boldmath $\alpha$}}

\usepackage{amssymb}

\journal{Physics Letters A}

\begin{document}

\begin{frontmatter}

\title{Gravitational lensing by spinning and radially moving lenses}
\author[federico,infn]{M. Sereno}
\ead{sereno@na.infn.it}

\address[federico]{Dipartimento di Scienze Fisiche, Universit\`{a} degli Studi di Napoli ``Federico II",
 Compl. Univ. Monte S. Angelo, via Cinthia, 80126, Napoli, Italia}
\address[infn]{Istituto Nazionale di Fisica Nucleare, Sez.
Napoli, Compl. Univ. Monte S. Angelo, Edificio G, 80126, Napoli,
Italia}

\begin{abstract}
The effect of currents of mass on bending of light rays is considered
in the weak field regime. Following Fermat's principle and the
standard theory of gravitational lensing, we derive the
gravito-magnetic correction to time delay function and deflection
angle caused by a geometrically-thin lens. The cases of both rotating
and shifting deflectors are discussed.
\end{abstract}

\begin{keyword}
Gravitational Lensing
\PACS 4.20.Cv \sep 4.25.Nx \sep 95.30.sf \sep 98.62.Sb
\end{keyword}

\end{frontmatter}

\section{Introduction}

Gravitational lensing has become a powerful tool for observational
astrophysics and cosmology. The bending of light has been widely
investigated in the framework of general relativity but some topics,
such as the dynamical properties of the deflectors, are usually
neglected. Peculiar and intrinsic motions of the lenses are expected
to be small second order effects. However, gravity induced by moving
matter is related to the dragging of inertial frames and the effects
of mass currents on the propagation of light signals deserve attention
from the theoretical point of view. Furthermore, since the impressive
development of technological capabilities, it is not a far hypothesis
to obtain observational evidences in a next future.

Lensing of light rays by stars with angular momentum has been
addressed by several authors with very different approaches. Epstein
\& Shapiro \cite{ep+sh80} performed a calculation based on the
post-newtonian expansion. Ib\'{a}\~{n}ez and coworkers \cite{iba83,ib+ma82}
resolved the motion equation for two spinning point-like particles,
when the spin and the mass of one of the particles were zero, by
expanding the Kerr metric in a power series of gravitational constant
$G$. Dymnikova \cite{dym86} studied the time delay of signal in a
gravitational field of a rotating body by integrating the null
geodesics of the Kerr metric. Glicenstein \cite{gli99} applied an
argument based on Fermat's principle to the Lense-Thirring metric to
study the lowest order effects of rotation of the deflector. The
listed results give a deep insight on some peculiar aspects of
spinning lenses but are very difficult to generalize. On the other
hand, Capozziello et al. \cite{cap+al99} discussed the
gravito-magnetic correction to the deflection angle caused by a
point-like, shifting lens in weak field regime and slow motion
approximation.

In this paper, we show that usual, basic assumptions of gravitational
lensing theory allow to consider effects of currents of mass in a very
general way. In the weak field approximation, we derive the
gravito-magnetic correction to the time delay function and the
deflection angle caused by a geometrically-thin lens. Our results,
based on Fermat's principle, apply to almost all observed
gravitational lensing systems and contribute to fill the gap between
the weak field approximation and a full theory of gravitational
lensing enlarged to any order of approximation.

\section{Gravito-magnetic correction}

We follow the standard hypotheses of gravitational lensing as
summarized in the monographs by Schneider et al. \cite{sef} and
Petters et al. \cite{pet+al01}. The gravitational lens is localized in
a very small region of the sky and its lensing effect is weak. The
deflector changes its position slowly with respect to the coordinate
system, i.e. the matter velocity is much less than $c$, the speed of
light; matter stresses are also small (the pressure is much smaller
than the energy density times $c^2$). In this weak field regime and
slow motion approximation, space-time is nearly flat near the lens. Up
to leading order in $c^{-3}$, the metric is
\beq
\lab{wf1}
ds^2 \simeq \left( 1+2\frac{\phi}{c^2}\right)c^2dt^2-8c dt \frac{ {\bf
V} {\cdot} d {\bf x}}{c^3}- \left( 1-2\frac{\phi}{c^2}\right)d{\bf x}^2;
\eeq
$\phi$ is the Newtonian potential,
\beq
\lab{wf2}
\phi (t, {\bf x}) \simeq \int_{\Re^3} \frac{\rho (t, {\bf x^{'}})}{ | {\bf x} -
{\bf x^{'}}|} d^3 x^{'};
\eeq
$\bf V$ is a vector potential taking into account the gravito-magnetic
field produced by mass currents,
\beq
\lab{wf3}
{\bf V} (t, {\bf x}) \simeq \int_{\Re^3} \frac{(\rho {\bf v})(t, {\bf
x^{'}})}{ | {\bf x} -{\bf x^{'}}|} d^3 x^{'},
\eeq
where ${\bf v}$ is the velocity field of the mass elements of the
deflector as measured in a coordinate system defined such that one of
its spatial coordinate axes follows the line from the observer to the
source, and the other two spatial axes lie in the plane of the sky. In
Eqs.~(\ref{wf2},\ref{wf3}), we have neglected the retardation
(Schneider et al. \cite{sef}). The line element in Eq.(\ref{wf1})
satisfies the weak-field condition once $|
\phi | \ll c^2$.

\subsection{The time delay function}

We assume that, during the time light rays take to traverse the lens,
the potentials in Eqs.~(\ref{wf2},\ref{wf3}) vary negligibly little.
Then, the lens can be treated as stationary. Under this hypothesis,
the space-time described by Eq.~(\ref{wf1}) can be interpreted as a
flat one with an effective index of refraction, $n$, given by
\cite{sef}
\beq
\lab{wf4}
n \simeq 1-2\frac{\phi}{c^2}+\frac{4}{c^3} {\bf V} {\cdot} {\bf e},
\eeq
where $\bf e$ is the unit tangent vector of a light ray. A light
signal emitted at the source will arrive after
\beq
\lab{wf5}
\1=\frac{1}{c}\int_p n \ ds=
\frac{1}{c}\int_p \left( 1-2\frac{\phi}{c^2}+\frac{4}{c^3} {\bf V} {\cdot} {\bf e} \right)\ ds,
\eeq
with the integral performed along the light trajectory $p$. The time
delay of the path $p$ relative to the unlensed ray $p_0$ is
\beq
\lab{wf6}
\1=\frac{1}{c}\left( \int_p n \ ds - \int_{p_0} n \ ds \right).
\eeq
Equation~(\ref{wf6}) can be expressed as a sum of geometrical and
potential time delays
\[
\1 =\1_{geom}+\1_{pot},
\]
with
\beq
\lab{wf7}
c\1_{geom}= \int_p \ ds - \int_{p_0} \ ds,
\eeq
due to the extra path length relative to the unperturbed ray, and
\beq
\lab{wf8}
c\1_{pot}= -\int_p \left( 2\frac{\phi}{c^2}-\frac{4}{c^3} {\bf V} {\cdot}
{\bf e} \right)\ ds,
\eeq
due to the retardation of the deflected ray caused by the
gravitational field of the lens.

In usual lensing phenomena, the deflection angle of a light ray is
very small. Furthermore, we treat the gravitational lens as thin. In
almost all astrophysical configurations, the extent of the lens in the
direction of the incoming ray is small compared to the distances
between lens and observer, $D_d$, and lens and source, $D_{ds}$, so
that the maximal deviation of the actual ray from a light path
propagating through unperturbed space-time is small compared to the
length scale on which the gravitational field changes. In this
picture, the deflection occurs essentially in a small region near the
deflector. The actual ray path can be approximated by combining its
incoming and outgoing asymptotes. This trajectory is a
piecewise-smooth null geodesic curve consisting of two null geodesics
of the unperturbed space-time, one from the source to the deflector
and one from the deflector to the observer. It is useful to employ the
spatial orthogonal coordinates $(\xi_1, \xi_2, l )$. The $l$-axis is
along the incoming light ray direction ${\bf e}_{in}$; the
$\xi_i$-axes lie in the plane of the sky. The lens plane corresponds
to $l=0$. With these assumptions, the geometrical time delay is
\cite{sef,pet+al01}
\beq
\lab{wf9}
c\1_{geom} \simeq \frac{1}{2}\frac{D_d D_{s}}{D_{ds}}\left|
\frac{\bfxi}{D_d}-\frac{\bfeta}{D_s} \right|^2,
\eeq
where $D_s$ is the distance from the observer to the source and
$\bfeta$ is the bidimensional vector position of the source from the
optical axis in the source plane.

The actual ray light is deflected, but if the deflection angle is
small, it can be approximated as a straight line in the neighbourhood
of the lens. This corresponds to the Born approximation, which allows
integrating along the unperturbed ray ${\bf e}_{in}$. We get
\beq
\lab{wf10}
c\1_{pot} \simeq -\frac{4
G}{c^2}\int_{\Re^2}d^2\xi^{'}\Sigma(\bfxi^{'})\left( 1-2\frac{\langle
{\bf v}{\cdot}{\bf e}_{in}\rangle_l(\bfxi^{'})}{c}\right){\rm ln}
\frac{|\bfxi -\bfxi^{'}|}{\xi_0}+\rm const.
\eeq
In Eq.~(\ref{wf10}), $\Sigma$ is the surface mass density of the
deflector,
\beq
\lab{wf11}
\Sigma(\bfxi)\equiv \int \rho(\bfxi,l)\ dl;
\eeq
$\langle {\bf v}{\cdot}{\bf e}_{in}\rangle_l$ is the weighted average, along
the line of sight, of the component of the velocity $\bf v$ orthogonal
to the lens plane,
\beq
\lab{wf12}
\langle {\bf v}{\cdot}{\bf e}_{in}\rangle_l (\bfxi)\equiv \frac{\int ({\bf v}(\bfxi,l){\cdot} {\bf e}_{in}) \ \rho(\bfxi,l)\ dl}{\Sigma(\bfxi)}.
\eeq
Adding the geometrical contribution, Eq.~(\ref{wf9}), and the
potential term, Eq.~(\ref{wf10}), we get the time delay of a
kinematically possible ray with impact parameter $\bfxi$ in the lens
plane, relative to the unlensed one for a single lens plane. The time
delay function is
\beq
\lab{wf13}
c\1= \frac{1}{2}\frac{D_d D_{s}}{D_{ds}}\left|
\frac{\bfxi}{D_d}-\frac{\bfeta}{D_s} \right|^2-\psi(\bfxi )+ \rm
const.,
\eeq
where $\psi$ is the deflection potential up to the order $v/c$,
\beq
\lab{wf14}
\psi(\bfxi ) \equiv \frac{4 G}{c^2}\int_{\Re^2}d^2\xi^{'}\Sigma(\bfxi^{'})\left( 1-2\frac{\langle {\bf v}{\cdot}{\bf e}_{in}\rangle_l(\bfxi^{'})}{c}\right){\rm ln} \frac{|\bfxi -\bfxi^{'}|}{\xi_0}.
\eeq
Hereafter, we will neglect the $const.$ in Eq.~(\ref{wf13}), since it
has no physical significance \cite{sef}. We remind that the time delay
function is not an observable, but the time delay between two actual
rays can be measured.

Since the potential time delay is a local effect which arises when a
ray traverses the neighbourhood of the lens, the time delay function
can be easily generalized to a cosmological context. In an homogeneous
and isotropic background perturbed by an isolated lens, $\1_{pot}$ has
to be redshifted; $\1_{geom}$, except for a cosmological factor
$(1+z_d)$ with $z_d$ the redshift of the deflector, takes the same
form of Eq.~(\ref{wf9}) once we consider the distances as angular
diameter distances \cite{sef,pet+al01}. The time delay measured at the
observer is
\beq
\lab{wf15}
c \1 =(1+z_d)\left\{ \frac{1}{2}\frac{D_d D_{s}}{D_{ds}}\left|
\frac{\bfxi}{D_d}-\frac{\bfeta}{D_s} \right|^2 - \psi(\bfxi ) \right\}.
\eeq
The velocity $\bf v$ is the peculiar velocity with respect to the
coordinate system. In a cosmological context, the recession velocity
of the deflector does not contribute to the gravito-magnetic
correction.

\subsection{The lens equation}

To derive the lens equation, we can apply Fermat's principle to the
bending of light caused by an isolated perturbation (Schneider et al.
\cite{sef} and Petters et al. \cite{pet+al01}). Fermat's principle
states that a light ray from a source to an observer follows a
trajectory, from among all kinematically possible paths, that is a
stationary value of the arrival path. In terms of a variational
formulation,
\[
\delta \int_p n \ ds =0,
\]
with the integral performed along the light trajectory. Actual light
rays, given the source position, are characterized by critical points
of $\1 (\bfxi)$, i.e. $\1 (\bfxi)$ is stationary with respect to
variations of $\bfxi$. The lens equation is then obtained calculating
\beq
\lab{wf16}
\nabla_{\bfxi}\1 (\bfxi)=0;
\eeq
we get
\beq
\lab{wf17}
\bfeta=\frac{D_s}{D_d}\bfxi-D_{ds}\bfal(\bfxi);
\eeq
$\bfal \equiv \nabla_{\bfxi}\psi$ is the deflection angle, i.e. the
difference of the initial and final ray direction. It is, to the order
$c^{-3}$,
\beq
\lab{wf18}
\bfal(\bfxi ) \equiv \frac{4 G}{c^2}\int_{\Re^2}d^2\xi^{'}\Sigma(\bfxi^{'})\left( 1-2\frac{\langle {\bf v}{\cdot}{\bf e}_{in}\rangle_l(\bfxi^{'})}{c}\right) \frac{\bfxi -\bfxi^{'}}{|\bfxi -\bfxi^{'}|^2}.
\eeq
In the thin lens approximation, the only components of the velocities
parallel to the line of sight enter the equations of gravitational
lensing. A change in position of the deflector orthogonal to the line
of sight can be noticeable in a variation of the luminosity of the
source but does not affect the individual light rays, i.e. does not
contribute to the gravito-magnetic correction.

For shifting lenses, $\langle {\bf v}{\cdot}{\bf e}_{in}\rangle_l (\bfxi )=
v_l$, the gravito-magnetic correction reduces to a multiplicative
factor to the zero order expressions. The deflection angle and the
related quantities, such as the optical depth, up to order $v/c$ are
derived from the zero-order expressions just by a product by $1
-2v_l/c$. For deflector moving towards the observer and far away from
the source ($v_l>0$), the optical depth decreases; for receding lenses
($v_l<0$), the deflection angle increases.

The velocity field of the deflector can break the axial symmetry of a
spherical distribution of mass with respect to the lensing effect. In
general, a slowly moving deflector, with surface mass density
$\Sigma^{\rm SLMO}$, has the same lensing effect of a really static
lens with
\beq
\lab{wf21}
\Sigma^{\rm STAT} (\bfxi ) = \Sigma^{\rm SLMO} (\bfxi ) \left( 1-2\frac{\langle {\bf v}{\cdot}{\bf e}_{in}\rangle_l(\bfxi )}{c}\right).
\eeq
As an illustrative case, let us consider a rotating lens, a classical
example of dragging of inertial frames. To simplify the calculations,
we will take the angular momentum of the deflector along the
$\xi_2$-axis and a constant angular velocity $\omega$. In this
configuration, the component of the velocity $\bf v$ orthogonal to the
lens plane is constant along the line of sight. Introducing polar
coordinates $(\theta, \xi)$ in the lens plane, we get
\[
\langle {\bf v}{\cdot}{\bf e}_{in}\rangle_l=-\omega \xi \cos \theta;
\]
for a general distribution of mass $\Sigma$, the deflection angles in
Eq.~(\ref{wf18}) reduces to
\beq
\lab{wf19}
\alpha_1 (\theta, \xi)= \frac{4 G}{c^2}\int_{0}^{\infty}\xi^{'}d \xi^{'}\int_{0}^{2 \pi}\left[ 1+2\frac{\omega}{c}\xi^{'}\cos \theta^{'}\right]\frac{\xi \cos \theta - \xi^{'}\cos \theta^{'}}{\xi^2+\xi^{'2}-2\xi \xi^{'}\cos (\theta^{'}-\theta)}\Sigma (\theta^{'},\xi^{'})d \theta^{'};
\eeq
\beq
\lab{wf20}
\alpha_2 (\theta, \xi)= \frac{4 G}{c^2}\int_{0}^{\infty}\xi^{'}d \xi^{'}\int_{0}^{2 \pi}\left[ 1+2\frac{\omega}{c}\xi^{'}\cos \theta^{'}\right]\frac{\xi \sin \theta - \xi^{'}\sin \theta^{'}}{\xi^2+\xi^{'2}-2\xi \xi^{'}\cos (\theta^{'}-\theta)}\Sigma (\theta^{'},\xi^{'})d \theta^{'}.
\eeq
For an axially-symmetric surface mass density, $\Sigma =\Sigma
(|\bfxi|)$, a not negligible value of $\omega /c$ breaks the symmetry
of the system and the full vector equation must be used.

\section{Conclusions}
We have considered the effects of gravity induced by currents of mass
by calculating the gravito-magnetic corrections to the time delay
function and the deflection angle for a single lens plane in the weak
field regime. Fermat's principle has been applied under the usual
assumptions of small deflection angles and geometrically thin lenses
to give simple formulae for a general mass distribution of the
deflector. For the point-like lens, our expression of the deflection
angle agrees with what found in Capozziello et al. \cite{cap+al99}.
Our analysis applies to almost all observed gravitational lensing
phenomena, both on galactic and extra-galactic scales.

We remark that a full analysis of higher order corrections to the
lensing theory makes possible a comparison between the predictions of
general relativity and extended relativistic theory of gravity where
corrections depend on the interaction scale \cite{cal+al00}. An
analysis to the lowest order might hide such differences.

In a forthcoming paper, we will discuss some specific lens models and
the perspective of precise observations to detect the gravito-magnetic
effect.

\end{document}